# Progress on FORS-Up: the first instrument using ELT technologies[*]

H. M. J. Boffin[a,†], V. Baldini[b], S. Bertocco[b], G. Calderone[b], R. Cirami[b], R.D. Conzelmann[a], I. Coretti[b], C. Cumani[a], D. Del Valle[a], F. Derie[a], P. A. Fuerte Rodríguez[a], P. Gutierrez Cheetham[a], J. Kosmalski[a], A. R. Manescau[a], P. Di Marcantonio[b], A. Modigliani[a], S. Moehler[a], C. Moins[a], D. Popovic[a], M. Porru[b], J. Reyes[a], R. Siebenmorgen[a], V. Strazzullo[b], and A. Sulich[b]

[a]ESO, Karl Schwarzschild str. 2, 85748, Garching bei München, Germany; [b]INAF – Osservatorio Astronomico di Trieste, via G.B. Tiepolo 11, Trieste, Italy

## ABSTRACT

ESO is in the process of upgrading one of the two FORS (FOcal Reducer/low dispersion Spectrograph) instruments – a multi-mode (imaging, polarimetry, long-slit, and multi-object spectroscopy) optical instrument mounted on the Cassegrain focus of Unit Telescope 1 of ESO's Very Large Telescope. FORS1 was moved from Chile to Trieste, and is undergoing complete refurbishment, including the exchange of all motorised parts. In addition, new software is developed, based on the Extremely Large Telescope Instrument Control Software Framework, as the upgraded FORS1 will be the first instrument in operations to use this framework. The new Teledyne e2V CCD has now been procured and is undergoing testing with the New Generation Controller at ESO. In addition, a new set of grisms have been developed, and a new set of filters will be purchased. A new internal calibration unit has been designed, making the operations more efficient.

**Keywords:** Very Large Telescope, ELT, FORS1, FORS2, focal reducer, instrumentation, upgrade, imaging, spectroscopy, polarimetry

## 1. THE "FORIPMOS" INSTRUMENTS

FORS1 and FORS2 are twin instruments, part of the first generation of ESO's Very Large Telescope' (VLT) suite of instruments. FORS stands for FOcal Reducer and low dispersion Spectrograph, but these instruments are really Swiss army's knives that are able to do much more: they do imaging (IMG; with pixel scales of 0.0625" or 0.125" – even if they are generally used in the latter mode and with a binning 2×2, giving a spatial resolution of 0.25"), imaging polarimetry (IPOL), long-slit (LSS) and multi-object spectroscopy (MOS, with 19 movable slitlets, or MXU, with masks), as well as spectro-polarimetry (PMOS). As such, these instruments should really be called FORIPMOS, for FOcal Reducer Imager, Polarimeter, and Multi-Object Spectrograph.

FORS1 was operational from 1999 to 2009, while FORS2 started operations in 2000 and hasn't stopped operating since. Together, they have been at the Cassegrain foci of all four Unit Telescopes of the VLT and have been among the most demanded instruments at this facility. Even today, FORS2 is still among the five most demanded instruments at the La Silla Paranal Observatory, with a request about three or four times higher than some of the newer but more specialised instruments. At the time of writing, both FORS instruments have led to the publication of more than 2,600 refereed papers, accounting for almost a fifth of all VLT-based papers. It is thus no surprise that it was decided to ensure that FORS will still be operational for the next 15 years. A project to upgrade this instrument (FORS-Up) has been started in 2020 as a collaboration between ESO and INAF – Astronomical Observatory of Trieste (INAF-OATS). It aims at bringing to the telescope in 2026 a refurbished instrument with a new scientific detector, an upgrade of the instrument control software and electronics, a new calibration unit, as well as a new set of filters and grisms, and a new software for preparing multi-object spectroscopic observations (FIMS). The new FORS will also serve as a test bench for the Extremely Large Telescope (ELT) standard technologies – among them the use of programmable logic controllers (PLCs) and of the features of the ELT Control Software. The project aims at minimising the downtime of the instrument by performing the upgrade on the

---

twin, currently decommissioned, instrument FORS1 and retrofitting the Mask Exchange Unit from FORS2 to FORS1. A presentation of the background to the project and the goals is given in [1] and in [2].

## 2. USING ELT TECHNOLOGIES

FORS1 was moved from La Silla to the INAF-OATS's integration hall at Basovizza (Fig. 1). There, the team successfully replaced the motors with new brushless Beckhoff motors in the three main sections of the instrument. In addition, for the control electronics, the current VME-based Local Control Unit (LCU) are being replaced, following the new standard adopted for the ELT instruments. The new control system is based on the Beckhoff TwinCAT, while keeping the same cabinet distribution as the current one. More details on this, as well as on the control of the MOS blades and the scientific exposure shutter, is available in [3, 4].

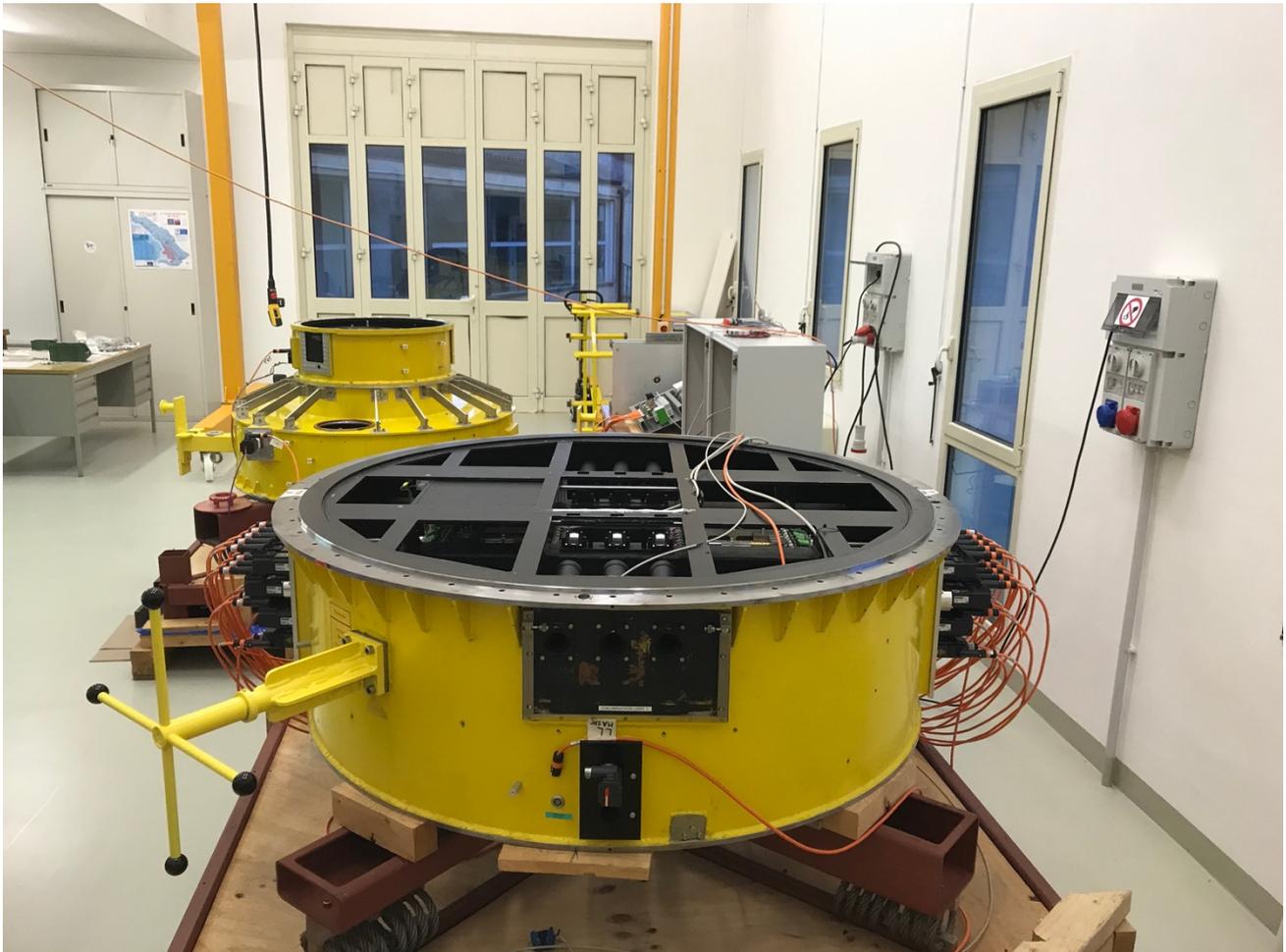

Figure 1. The dismounted FORS1 in the integration hall of INAF-OATS. In the front is the MOS section, which has been fully populated with 38 new motors. The back shows the Camera and Filter wheel section of the instrument.

Similarly to the control electronics, the control software is being rewritten, following ELT standards. Thus, the VLT-based control software is reimplemented within the ELT Instrument Control Software Framework. A new set of Graphical User Interfaces is also under development, making use of the Control User Interface Toolkit provided by ESO (Fig. 2). More details are available in [5]. Because FORS1 will be using the ELT framework, ahead of any other instruments, and at the

VLT, there is a need for a dedicated interface, which is provided by ESO: the ELT-VLT Gateway, that forwards telescope commands from the Instrument Workstation (IWS) to the Telescope Control System (TCS) and returns the telescope status and data products from the TCS to the IWS.

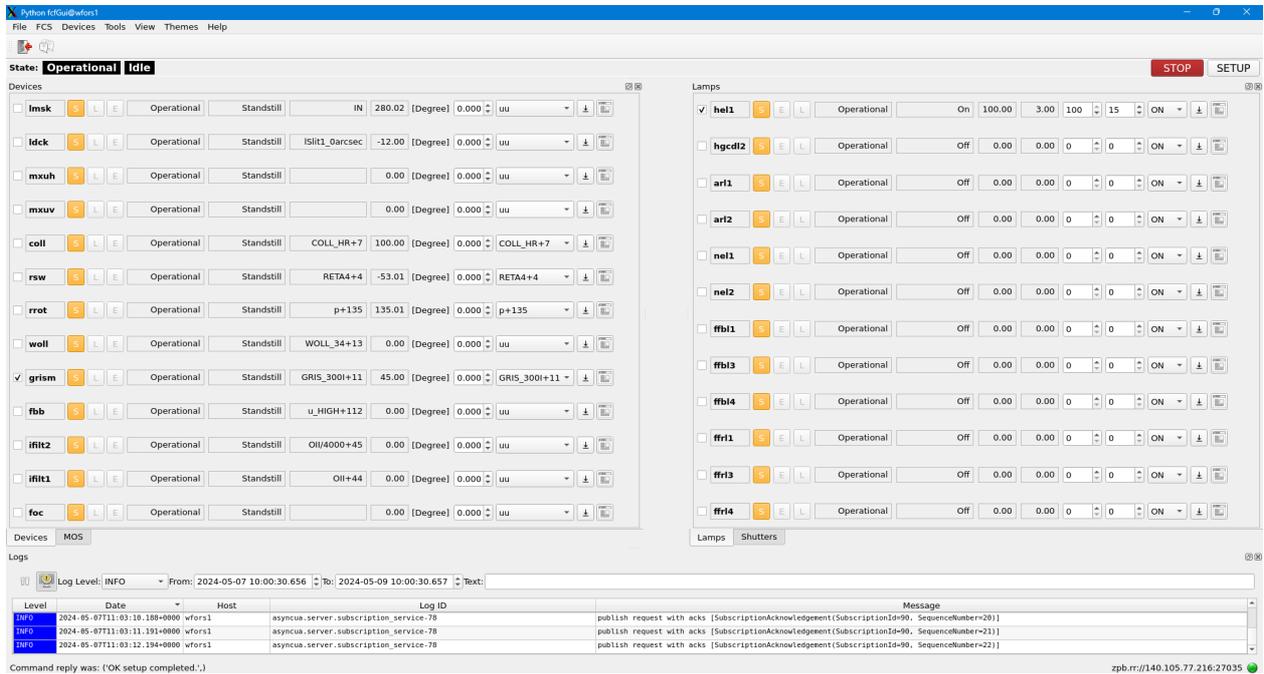

Figure 2. The current design of the new FORS Control System (formerly the ICS).

## 3. OTHER ACHIEVEMENTS

### 3.1 New detector

One of the main goals of FORS-Up foresees replacing the two (red and blue) science detectors that are each composed of two 2k × 4k chips, with a single 4k × 4k chip with an excellent response in both the blue and red parts of the spectrum. The proposed chip is a variant of the CCD used in the VLT Multi Unit Spectroscopic Explorer (MUSE) instrument – the Teledyne e2v CCD231-84 – with fringe suppression technology and an enhanced anti-reflection coating. Most importantly, the current FORS2 cryostat and optical mount are compatible with the selected new detector, which simplifies the design, development, and operation of the new FORS. Such a detector has now been delivered to ESO, where it is undergoing extensive testing and qualification (Fig. 3).

The new FORS1 detector will be named Nonino, in honour of our former colleague. It will be managed by the New General Controller II (NGC-II) via a dedicated workstation. The detector can be read via 1- or 4-ports, leading to read-out times between 6 seconds (2×2 binning, 4-ports, fast mode) to 179 seconds (1×1 binning, 1-port, science mode). These modes need to be further qualified in terms of read-out noise, gain, and dark current, to allow for a final decision on which modes will be offered during operations.

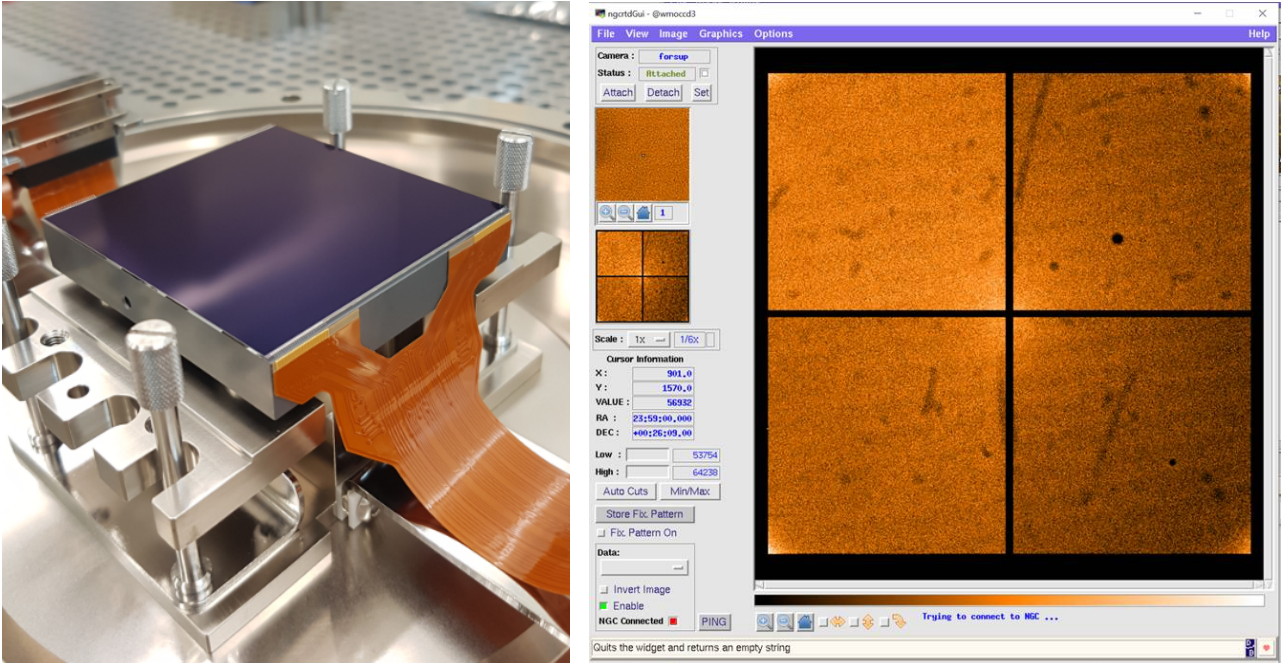

Figure 3. The new FORS1 detector on its test-bench at ESO (left) and a raw flat field image (right).

### 3.2 FORS1 Carriage

The Cassegrain carriage of FORS1, which is used for all major transport and handling operations, needed to be reconditioned and recertified for further operations. This was done by MBE1 Maschinenbau Eggerstorfer GmbH, the German company that originally built the carriage, and the carriage was redelivered to ESO, almost as new.

### 3.3 Filters and grisms

In the Phase A of the project, it was identified that the project should also ensure that the newly furbished FORS would have an ideal suite of filters and grisms.

There is already a very extensive suite of such optical components, in use at FORS2. Still, a dedicated study has identified the usefulness of procuring a new set of filters for FORS1 to complement the best ones we already have, in particular with an eye to the existing and forthcoming sky surveys, such as SDSS (Sloan Digital Sky Survey), Pan-STARRS, DES (Dark Energy Survey), and Vera Rubin Observatory's Legacy Survey of Space and Time (LSST). Based on this, four additional filers, similar to the ULTRACAM *r* and *i* filters (based on the SDSS set), and LSST-like *z* and *y* filters, will be added to the original filter set, as an ideal complement to the current FORS2 $u_{HIGH}$ and $g_{HIGH}$ filters for the bluest bands (Fig. 4). Having filters close to the LSST ones is very useful as all FORS1 pointings will be within the LSST footprint. Once the colour terms between the two systems will be determined on sky, this will result in an almost immediate photometric calibration for FORS1 observations in the imaging mode. The ULTRACAM *r* and *i* filters, already in use at ESO (at the NTT in La Silla), have been selected for their very high transmission, as well as slightly better short and long wavelength cutoff with respect to their LSST counterparts.

Concerning grisms, FORS2 has 15 grisms available, with spectral resolution (1" slit) between 260 and 2600, and covering the whole FORS2 wavelength range, from 330 to 1100 nm. The list of grisms is available at [6]. A dedicated study has highlighted the usefulness of having three new holographic grisms (VPHG) for FORS1. One, with a spectral resolution of about 800, covering the 330 to 620 nm range, but with very high transmission, to replace the current 600B grism. In addition, two higher resolution grisms centred on, respectively, the sodium (580 nm) and the potassium (770 nm) lines would be extremely useful to perform with even better efficiency transmission spectroscopy of exoplanets [7-10]. The three VPHGs have been designed by means of a rigorous coupled-wave analysis approach and are presented in [11].

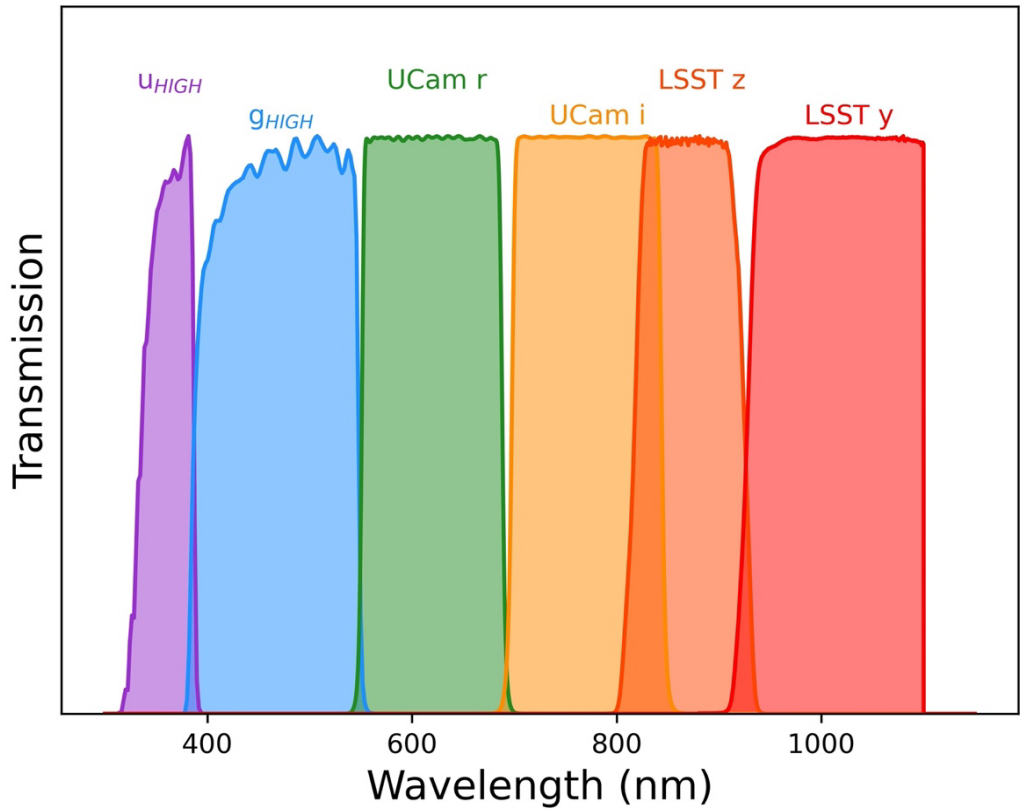

Figure 4. The four new FORS1 filters will ideally complement the current $u_{HIGH}$ and $g_{HIGH}$ filters to cover the whole FORS1 wavelength range. Many more filters, including narrow-band ones, are also available.

## 4. THE INTERNAL CALIBRATION UNIT

For calibrating the data, FORS2 currently uses an internal and an external calibration unit (CU), hosting the arc and flat field lamps. The external calibration unit, which is used for all spectroscopic and spectrophotometric calibrations, is unfortunately rather unreliable. One of the major drawbacks of the actual external calibration unit is that the calibration light needs to be projected from outside of the instrument, which puts many limitations on the calibration process and the regular activities in the telescope – it is, for example, not possible to take calibrations with the telescope at a position other than at zenith, which leads then to shifts in the wavelength solution. In addition, the configuration is quite inefficient in terms of the injected light for calibration purposes, requiring a rather high intensity from the light source to correctly illuminate the instrument's detector.

As part of FORS-Up, it is now foreseen to put again all calibration lamps in an internal calibration unit, merging the external with the (original) internal CUs, and providing the calibration light required for the instrument calibration for all spectroscopic modes (LSS, MOS, MXU, PMOS). The new CU is not designed for imaging modes (IMG, IPOL) of the FORS1 instrument, as twilight flats are used instead.

The new FORS1 CU shall address the main issues noted in the operation of the existing calibration units:

- The lamp used for flat fields in the blue range is a halogen lamp located in the external CU. To guarantee the right illumination in the UV, it is a high-power lamp (250W), with a rather high dissipation. This lamp fails regularly and replacing it is quite time consuming, as aside the bulb replacement, it is also necessary to fine tune the diaphragm and the lamp's position to keep the calibration flux within the operational thresholds.

- One will also need to include an automatic power off for the lamps when the calibration templates fail for whatever reason. Currently, the calibration lamps remain on whenever there is any problem during the execution of the calibration procedures, until someone powers it off manually. This causes an early demise of the lamps.

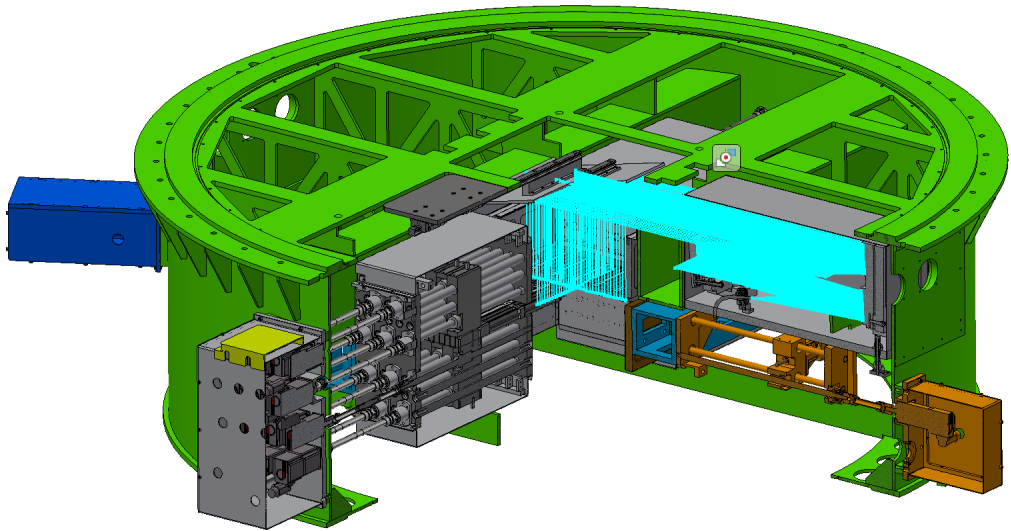

Figure 5. CAD model of the FORS1 top section, with the new calibration unit projection optics. The cyan light indicates the light path from the lightguide to the instrument entrance focal plane.

For the new CU, it has been decided to adopt as baseline for the lamp modules and control electronics, the concept developed for MUSE, where the light sources are managed in individual modules. The use of new standards for the electronics and control system shall enable not only to control the individual light sources, but to gather detailed performance monitoring to ensure the quality of the calibration light. This information is crucial for preventive maintenance, and to guarantee the quality of the calibration sources for standard operation. Lamp modules and associated control electronics will be enclosed in a dedicated cabinet that will be attached to the instrument body together with the other three new control electronic cabinets. The light generated in the lamp modules is delivered to the projection optics module through lightguides, as in MUSE. The projection optics has been designed to fit in the FORS1 top section (Fig. 5). This subsystem projects the light at the instrument entrance focal plane and is delivered to the detector by the FORS1 optical system in each of its operational configurations (Fig. 6).

The new compact design will in the end provide between 4 to 10 times higher flux than the current CUs, while using lamps of 50W maximum. This will considerably reduce the time needed to calibrate the data, ensure extended longevity of the lamps, ease the maintenance, and allow calibration to be done at any position of the telescope.

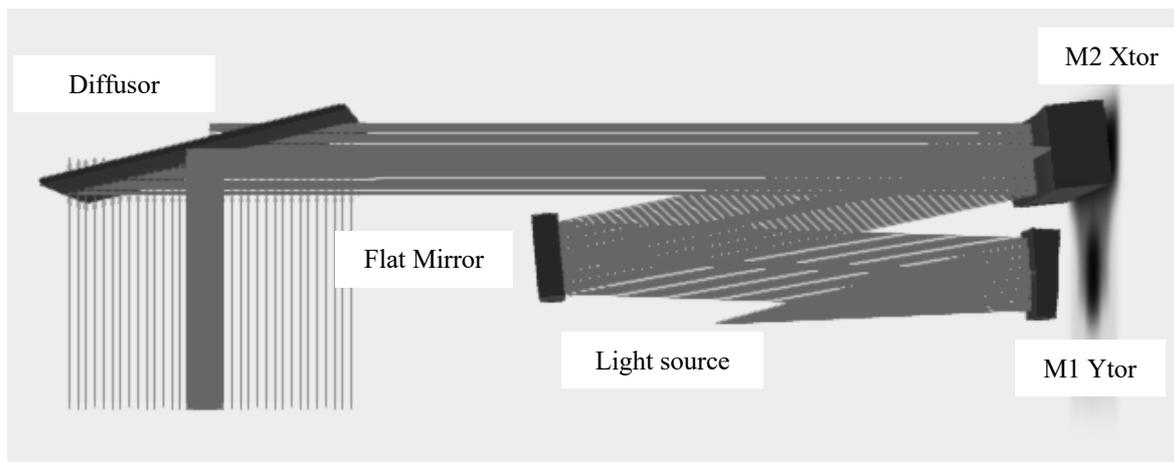

Figure 6. Raytracing in the proposed projection system for the new FORS1 Internal Calibration Unit.